\documentclass[aps,pra,twocolumn,showpacs,amsmath,amssymb,superscriptaddress]{revtex4-1}
\usepackage{graphicx}
\begin{document}
\title{Medium effects close to $s$- and $p$-wave Feshbach resonances in atomic Fermi gases}
\author{Renyuan Liao}
\affiliation{College of Physics and Energy, Fujian Normal University, Fuzhou 350108, China}
\affiliation{National Laboratory for Condensed Matter Physics, Institute of Physics, Chinese Academy of Sciences, Beijing 100190, China}
\author{Khandker F. Quader}
\affiliation{Physics Department, Kent State University, Kent, OH 44240, USA}
\date{\today}

\begin{abstract}
Many-body effects may influence properties, such as scattering
parameters, nature of pairing, etc., close to a Feshbach resonance in
the fermion BEC-BCS crossover problem. We study effects such as
these using a tractable crossing-symmetric approach. This method
allow us to include quantum fluctuations, such as, density,
current, spin, spin-current and the higher-order fluctuations in a
self-consistent fashion. The underlying fermion interaction is
reflected in the ``driving" term. We perform calculations here on both
Bose-Einstein condensate (BEC) and BCS sides, and taking the driving term to be
finite range, and of arbitrary strength. These are related to two-body
singlet and triplet scattering parameters, and can be connected with
experimental $s-$ and $p$-wave Feshbach resonances. We include the $\ell=0$
density and spin fluctuations, as well as $\ell=1$ current and
spin-current fluctuations. We calculate renormalized scattering
amplitudes, pairing amplitudes, nature of pairing, etc., on both the
BEC and BCS sides. We then compare our results qualitatively with
experiments.
\end{abstract}
\maketitle

\section{Introduction}
The impact of ultracold atomic and molecular quantum gases on present-day physics is associated with
 the extraordinary degree of control that such systems offer to investigate the fundamental behavior of quantum matter under various conditions~\cite{CHI10}. Recent experimental achievements in the field of ultracold Fermi gases are based mainly on the possibility of tuning the scattering length $a_s$, in particular to values much large than the mean interatomic distance, by changing an external magnetic field~\cite{GIO08}. This situation exists near the so-called Feshbach resonances.

Resonances in general refer to the energy-dependent enhancement of
interparticle scattering cross section due to the existence of
a metastable state. For Feshbach resonances, the metastable
state is described in terms of coupling of a bound state of a
subsystem to its environment. By tuning a magnetic field, it is
possible to obtain a quasidegeneracy between the relative energy of
two colliding atoms and that of a weakly bound molecular state. As
the quasibound state passes through a threshold, the scattering length
can be varied, in principle, from positive to negative infinity. The Feshbach
resonances were observed in bosons~\cite{Inouye98,Courteille98,Roberts98,Marte02},
in fermions between distinct spin states~\cite{Loftus02,Hara02,Dieckmann02}, and
in a single-component Fermi gas~\cite{Regal032}. In this manner the
interactions between the atoms can be strongly enhanced by an
external magnetic bias field, giving rise to the BEC-BCS crossover
phenomena~\cite{Eagles69,Leggett80,Nozieres85}. As a result of the
atomic physics of the Feshbach resonance, the nature of the Cooper
pairs in the BEC-BCS crossover is, however, not solely determined
by the interaction strength or scattering length but, in principle,
also depends on the width of the Feshbach resonance. In the limit of
an infinitely broad resonance, the properties of the gases can be
described by a single-channel theory that requires only the resonant
scattering length as an experimental input. In general, however, a
two-channel theory is needed. This is, in particular, true for the
description of the wave function of the Cooper pairs that plays an
important role in the BEC-BCS crossover.

   Near a resonance, fluctuations can be quite important since the system
is no longer in the dilute limit, and therefore one may expect
contributions from higher angular momenta and from quantum fluctuations or influence of collective modes. Generally, inclusion of
many-body effects in these systems is not easy, and one usually stops at the level of random phase approximation (RPA).

 In this paper, we aim at presenting a different scenario which
emphasizes the role of the exchange particle-hole fluctuations in addition to RPA in
the direct particle-hole channel. Our calculation is based on the induced interaction model of Babu and
Brown~\cite{Babu73,Quader87}, subsequently generalized and termed as the ``crossing-symmetric approach"~\cite{Ainsworth83}. This takes into account a properly
antisymmetrized, effective two-body interaction that reproduces the
correct low-energy physics and also suppresses any spurious ground
states. The physics is described in terms of the Landau interaction parameters and
scattering amplitudes to be discussed. In a previous study  Gaudio {\it{et al.}}~\cite{GAU07} considered $s$-wave scattering by including $\ell=0$ density and spin fluctuation. We shall study $s$- and $p$- wave scattering and include both $\ell=0$ density and spin fluctuations and $\ell=1$ current and spin-spin fluctuations.
\section{Theoretical approach}
\subsection{The crossing-symmetric method}
The crossing-symmetric method~\cite{Babu73,Ainsworth83,Quader87}
was formulated to calculate the effective quasiparticle
interactions in Fermi systems. Due to an appropriate compromise
between microscopic and phenomenological approaches, it has been
successfully applied to a number of Fermi systems: liquid
$^3$He~\cite{Babu73,Ainsworth83,Quader85}, $^3$He-$^4$He
mixtures~\cite{Yi83}, paramagnetic
metals~\cite{Ainsworth83}, heavy-fermion systems~\cite{Bedell85},
nuclear matter~\cite{Sjoberg73}, and ultracold atomic Fermi gases~\cite{GAU07}. It has been
known~\cite{Babu73,Bedell83,Bedell84,Bedell85} that a consistent
Fermi-liquid theory cannot be formulated in terms of short-range
effective interactions alone; collective excitations generated by
these must be exchanged between quasiparticles.\\

  The main point is that the contributions to Landau interaction
 $f_{\mathbf{p}\mathbf{p}^\prime}^{\sigma\sigma^\prime}$ can be separated into two parts~\cite{Quader87}:
 \begin{equation}
  f_{\mathbf{p}\mathbf{p}^\prime}^{\sigma\sigma^\prime}=d_{\mathbf{p}\mathbf{p}^\prime}^{\sigma\sigma^\prime}+
  I_{\mathbf{p}\mathbf{p}^\prime}^{\sigma\sigma^\prime}\left[f_{\mathbf{p}\mathbf{p}^\prime}^{\sigma\sigma^\prime}\right],
 \end{equation}
 where the induced part $I_{\mathbf{p}\mathbf{p}^\prime}^{\sigma\sigma^\prime}$, a function of
 the Landau interactions $f_{\mathbf{p}\mathbf{p}^\prime}^{\sigma\sigma^\prime}$ themselves, is
 particle-hole reducible in the exchange particle-hole ($u$) channel, whereas the direct part $d_{\mathbf{p}\mathbf{p}^\prime}^{\sigma\sigma^\prime}$ is not
 particle-hole reducible in either the direct particle-hole ($t$) channel or the crossed particle-hole ($u$) channel. It is important to note that the direct interaction is
 model dependent, as it gives information about the underlying Hamiltonian, and that the induced interaction is
 a purely quantum effect, arising from the exchange diagrams required to antisymmetrize the
 effective two-body scattering amplitude. \\

   For a two-component fermionic system, the Landau interaction can be expressed as
\begin{equation}
   F_{\mathbf{p}\mathbf{p}^\prime}^{\sigma\sigma^\prime}=F_{\mathbf{p}\mathbf{p}^\prime}^{s}+
   F_{\mathbf{p}\mathbf{p}^\prime}^{a}\vec{\sigma}\cdot\vec{\sigma}^\prime,
\end{equation}
where $F_{\mathbf{p}\mathbf{p}^\prime}^{s}$ and $ F_{\mathbf{p}\mathbf{p}^\prime}^{a}$ can be related to induced-interaction equations as follows~\cite{Quader87}:
\begin{eqnarray}
  &&F_{\mathbf{p}\mathbf{p}^\prime}^{s}=D_{\mathbf{p}\mathbf{p}^\prime}^{s}+\frac{1}{2}\frac{F_0^sU_0(q')F_0^s}{1+F_0^sU_0(q')}+\frac{3}{2}\frac{F_0^aU_0(q')F_0^a}{1+F_0^aU_0(q')}\nonumber\\
             &
             &+\frac{1}{2}\left[1-\frac{q'^2}{4k_F^2}\right]\left[\frac{F_1^sU_1(q')F_1^s}{1+F_1^sU_1(q')}+3\frac{F_1^aU_1(q')F_1^a}{1+F_1^aU_1(q')}
             \right]\label{eq:Fs},
\end{eqnarray}
\begin{eqnarray}
  &&F_{\mathbf{p}\mathbf{p}^\prime}^{a}=D_{\mathbf{p}\mathbf{p}^\prime}^{a}+\frac{1}{2}\frac{F_0^sU_0(q')F_0^s}{1+F_0^sU_0(q')}-\frac{1}{2}\frac{F_0^aU_0(q')F_0^a}{1+F_0^aU_0(q')}\nonumber\\
             &&+\frac{1}{2}\left[1-\frac{q'^2}{4k_F^2}\right]\left[\frac{F_1^sU_1(q')F_1^s}{1+F_1^sU_1(q')}-\frac{F_1^aU_1(q')F_1^a}{1+F_1^aU_1(q')}
             \right]\label{eq:Fa}.
\end{eqnarray}
The momentum transfer in the crossed particle-hole
channel is $q^{\prime}=|\mathbf{p}-\mathbf{p}^\prime|=k_F\sqrt{1-\cos{\theta_L}}$, with the Landau angle
$\theta_L=\cos^{-1}{(\mathbf{p}\cdot\mathbf{p}^\prime)}$. $U_0(q^\prime)$ and $U_1(q^\prime)$ are the
Lindhard functions or density-density and current-current
correlation functions, respectively.
The first term in
Eqs.~(\ref{eq:Fs}) and ~(\ref{eq:Fa}) is the so-called direct
interaction. The direct term is designed to convey the fact that two
quasi-particles can directly scatter via some effective potential
and repeatedly so , as in a $T$-matrix. It is of short range and
contains information about the underlying Hamiltonian of the system
under consideration. Thus, it is the ``driving" term. The induced term
is of somewhat longer range since two particles can scatter via an
interaction mediated by another particle. Equations.~(\ref{eq:Fs}) and (\ref{eq:Fa}) are nonlinear coupled
equations. To solve these, we need to do Legendre projections:
\begin{eqnarray}
     F_{\mathbf{p}\mathbf{p}^\prime}^{s,a}=\sum_{l}F_{l}^{s,a}P_l(cos\theta_L),\\
     D_{\mathbf{p}\mathbf{p}^\prime}^{s,a}=\sum_{l}D_{l}^{s,a}P_l(cos\theta_L).
\end{eqnarray}
 Ainsworth {\it{et al.}}~\cite{Ainsworth83} treated $D_0^s$, $D_0^a$, and $D_1^s$
 phenomenologically so as to reproduce the empirical Landau
 parameters $F_0^s$, $F_0^a$, $F_1^s$, and $F_1^a$ and predicted
 the higher-order $F_{l}^{s,a}$'s ($l\geq$1). Then effective pairing interaction can
 be obtained~\cite{PAT75}:
 \begin{eqnarray}
    g_s^{eff}&=&\left[A_0^s-3A_0^a-(A_1^s-3A_1^a)\right]/4=A_s/4,\\
    g_t^{eff}&=&\left[A_0^s+A_0^a-(A_1^s+A_1^a)\right]/12=A_t/12,
 \end{eqnarray}
 where
 \begin{eqnarray}
   A_{l}^{s,a}= \frac{F_l^{s,a}}{1+F_{l}^{s,a}/(2l+1)}
 \end{eqnarray}
    is the $l$-partial wave symmetric (s), antisymmetric (a) scattering amplitude and $A_s$ and $A_t$ are the pairing-channel singlet and triplet scattering amplitude respectively. The vanishing of forward scattering of two particles of equal spin yields the Landau sum rule $\sum_l(A_l^s+A_l^a)=0$, which provides a test for obtained Landau parameters.

\subsection{The driving term near a Feshbach resonance }
In the crossing-symmetric approach, the form of the direct interaction used to derive the induced
interactions must be determined. In general, it represents the sum of
all particle hole irreducible interactions. A self-consistent calculation
could be performed starting with a $T$-matrix direct interaction if a
more general interaction were used to derive the $T$-matrix
calculation. According to the proposal by Bedell and
Ainsworth~\cite{Ainsworth87}, the direct interaction is the Fourier
transform of an effective quasi-particle potential. From this
potential the quasiparticle scattering amplitude $f_\mathbf{k}(\phi)$ is
given by
\begin{equation}
     f_{\mathbf{k}}(\phi)=\frac{-m^*}{4\pi}\int e^{i\mathbf{q}\cdot\mathbf{r}}V_{eff}(\mathbf{r},\mathbf{k})d^3\mathbf{r}
     \label{eq:sa},
\end{equation}
where $\hbar=1$, $q^2=|\mathbf{k}-\mathbf{k}^\prime|^2=2k^2(1-\cos{\phi})$,
$2k^2=k_F^2(1-\cos{\theta})$ and the quasiparticle mass $m^*=m(1+F_1^s/3)$. The relative momentum of the
incoming (outgoing) particles is $\mathbf{k}$($\mathbf{k}^\prime$), and the angle between the incoming and scattering
plane is $\phi$. Equation~(\ref{eq:sa}) is restricted to the Fermi
surface; therefore $f_\mathbf{k}(\phi)$ depends on only two variables,
$\theta$ and $\phi$. According to the effective range expansion, the
effective potential $V_{eff}(\mathbf{r},\mathbf{k})$ is, in general, nonlocal and can
be expanded in powers of $\mathbf{k}^2$,
\begin{equation}
   V_{eff}(\mathbf{r},\mathbf{k})=U(\mathbf{r})+\frac{1}{3}\mathbf{k}^2\mathbf{r}^2W(\mathbf{r})+...,
\end{equation}
where $U(\mathbf{r})$ and $W(\mathbf{r})$ are local potentials. Keeping order of $\mathbf{k}$ to 2
yields a three-parameter approximation to $f_\mathbf{k}(\phi)$,
\begin{equation}
  f_{\mathbf{k}}(\phi)\simeq\frac{m^*}{m}\left[-a_s+3k^2a_t(1-cos\phi)-3\mathbf{k}^2b_t\right],
  \label{eq:sca}
\end{equation}
where $a_s=m\int \mathbf{r}^2U(\mathbf{r})d^3\mathbf{r}$, $a_t=(m/9)\int \mathbf{r}^4U(\mathbf{r})d^3\mathbf{r}$, and
$b_t=(m/9)\int r^4 W(\mathbf{r})d^3\mathbf{r}$. As pointed out in Ref.~\cite{Ainsworth87}, the triplet quasiparticle scattering volume $a_t$ and the nonlocal part of the effective potential $b_t$ are $p$ wave in nature and thus sample relatively little of the repulsive core of the bare interaction. Therefore they should not have a strong density dependence. The $a_t$ is a finite-range correction to the contact interaction; this allows some interaction between particles of the same spin. The nonlocal piece of the direct interaction results from the coupling of quasiparticle currents. The
direct interaction for particles of parallel spin is
\begin{eqnarray}
    d^{\uparrow\uparrow}(\theta,\phi)&=&-\frac{4\pi}{m^*}\left[f_\mathbf{k}(\phi)-f_\mathbf{k}(\phi+\pi)\right]\nonumber\\
                                     &=&\frac{12\pi}{m}k_{F}^2a_t(1-\cos{\theta})\cos{\phi}.
\end{eqnarray}
Similarly for particles of opposite spin,
\begin{eqnarray}
    d^{\uparrow\downarrow}(\theta,\phi)&=&-\frac{4\pi}{m^*}f_\mathbf{k}(\phi)\nonumber\\
                                       &=&\frac{4\pi}{m}[a_s-\frac{3}{2}k_F^2a_t(1-\cos{\theta})(1-\cos{\phi})\nonumber\\
                                       &&+\frac{3}{2}k_F^2b_t(1-\cos{\theta})].
\end{eqnarray}
The standard Landau parameters are the $\mathbf{q}=0$ values of momentum
dependent functions $F_l^{s,a}(\mathbf{q})$. In the limit of
$\mathbf{q}\rightarrow0$, $\cos{\theta}=\cos{\theta_L}$, $\cos{\phi}=1$. $d^{\uparrow\uparrow}$, $d^{\uparrow\downarrow}$ could be rewritten as
\begin{eqnarray*}
  d_{\mathbf{p}\mathbf{p}^\prime}^{\uparrow\uparrow}&=&d_{0}^{\uparrow\uparrow}+d_{1}^{\uparrow\uparrow}P_1(\hat{\mathbf{p}}\cdot\hat{\mathbf{p}^\prime})=\frac{12\pi}{m}k_F^2a_t(1-\cos{\theta_L}),\\
  d_{\mathbf{p}\mathbf{p}^\prime}^{\uparrow\downarrow}&=&d_{0}^{\uparrow\downarrow}+d_{1}^{\uparrow\downarrow}P_1(\hat{\mathbf{p}}\cdot\hat{\mathbf{p}^\prime})=\frac{4\pi}{m}[a_s+\frac{3}{2}k_F^2b_t(1-\cos{\theta_L})].\\
\end{eqnarray*}
Solving the above equations, we obtain
\begin{eqnarray}
  d_0^{\uparrow\uparrow}&=&\frac{12\pi}{m}k_F^2 a_t,\nonumber \\
  d_0^{\uparrow\downarrow}&=&\frac{4\pi}{m}a_s+\frac{6\pi}{m}k_F^2 b_t,\nonumber\\
  d_1^{\uparrow\uparrow}&=&-\frac{12\pi}{m}k_F^2 a_t,\nonumber\\
  d_1^{\uparrow\downarrow}&=&-\frac{6\pi}{m}k_F^2 b_t.
\end{eqnarray}
The direct interactions $D^s$ and $D^a$ are linear combinations of
$d^{\uparrow\uparrow}$ and $d^{\uparrow\downarrow}$:
\begin{eqnarray}
 D^s&=&\frac{N(0)}{2}(d^{\uparrow\uparrow}+d^{\uparrow\downarrow}),\\
 D^a&=&\frac{N(0)}{2}(d^{\uparrow\uparrow}-d^{\uparrow\downarrow}),
\end{eqnarray}
where $N(0)=k_Fm^*/\pi^2$ is the density of states at Fermi surface.
 For a pure $s$-wave resonance, we set $a_t=0 , b_t=0$:
\begin{eqnarray}
 D_0^s=\frac{2\pi}{m}a_s N(0)=U_0/2, D_1^s=0\\
 D_0^a=-\frac{2\pi}{m}a_s N(0)=-U_0/2, D_1^a=0.
\end{eqnarray}
For a pure $p$-wave resonance, we set $a_s=0$:
\begin{eqnarray}
   D_0^s=\frac{3\pi k_F^2}{m}(2a_t+b_t)N(0), D_1^s=-D_0^s;\\
   D_0^a=\frac{2\pi k_F^2}{m}(2a_t-b_t)N(0), D_1^a=-D_0^a.
\end{eqnarray}
\subsection{Connection to the usual scattering parameters}
The two-body scattering amplitude is
\begin{eqnarray}
 f(\theta)&=&\frac{1}{2ik}\sum_{l=0}^{\infty}(2l+1)(e^{2i\eta_l}-1)P_l(\cos{\theta})\nonumber\\
          &=&\sum_{l=0}^{\infty}(2l+1)f_l(k)P_l(\cos{\theta}),
          \label{eq:c1}
\end{eqnarray}
 where the \emph{l}th angular momentum channel is given
 by~\cite{Landau94}
\begin{equation}
   f_l(k)=\frac{k^{2l}}{-a_l^{-1}+r_l k^2-ik^{2l+1}}.
   \label{eq:c2}
\end{equation}
According to the effective range expansion, we have
\begin{equation}
 k^{2l+1}\cot{\eta_l}=-\frac{1}{a_l}+r_l k^2+....
 \label{eq:c3}
\end{equation}
From the induced interaction description, as
$q=0$, $2k^2=k_F^2(1-\cos{\theta)}$. Equation~(\ref{eq:sca}) becomes
\begin{equation}
  f_\mathbf{k}(\theta)=\frac{m^*}{m}[-a_s-\frac{3}{2}k_F^2 b_t+\frac{3}{2}k_F^2 b_t\cos{\theta}]
  \label{eq:c4}
\end{equation}
Combining Eqs.~(\ref{eq:c1})-(\ref{eq:c4}), we
reach
\begin{eqnarray}
   \frac{m^*}{m  }\frac{1}{2}k_F^2
   b_t=\left|\frac{k^2}{-a_1^{-1}+r_1k^2-ik^3}\right|=\frac{k^2}{\sqrt{(k^3\cot{\eta_1})^2+k^6}}.\nonumber\\
\end{eqnarray}
\begin{figure}
\centering
\includegraphics[scale=0.33]{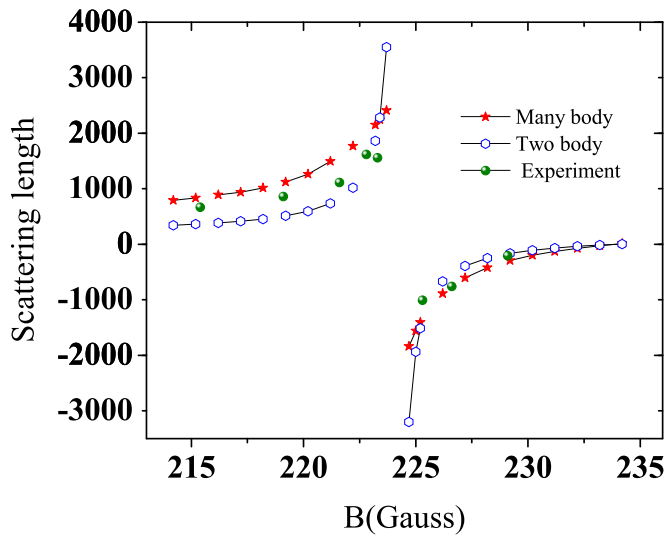}
\caption{(Color online) The $s$-wave scattering length as a function of the magnetic
field $B$ on both sides of the Feshbach resonance in $^{40}z$K, using data
from Regal and Jin~\cite{Regal03}. The density here is
$n=5.8\times10^{13}cm^{-3}$, with $\Delta B=9.7G$, and the Feshbach
resonance occurs at $B_0=224.21G$. The scattering length is measured in terms of the Bohr radius. }\label{fig:graph1}
\end{figure}

\begin{figure}
\centering
\includegraphics[scale=0.33]{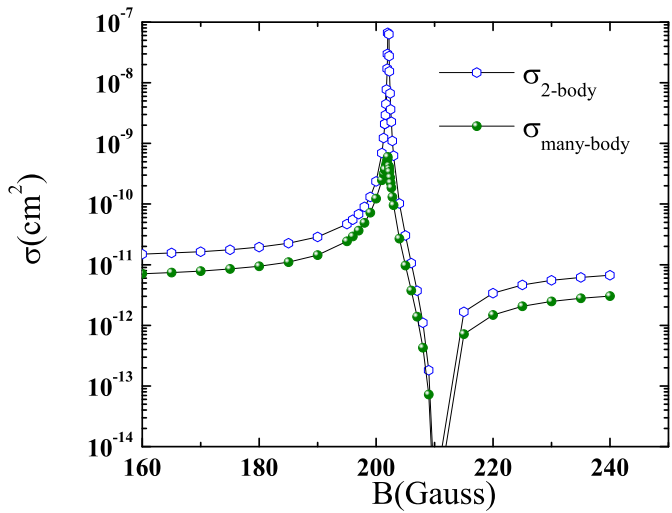}
\caption{(Color online) The scattering cross section of a $s$-wave resonance between
$^{40}$K atoms in $|f=9/2,m_f=-9/2>$ and $|f=9/2,m_f=-7/2>$
states. The data used~\cite{Regal032} for calculation are as follows: number density
is $n_{pk}=1.5\times10^{13}cm^{-3}$, with $\Delta B=7.8G$, and
the Feshbach resonance occurs at $B_0=202.1G$.}\label{fig:graph2}
\end{figure}
\section{Calculations and Results}
\subsection{Close to an $s$-wave resonance}

In the vicinity of a Feshbach resonance, the $s$-wave scattering
length $a(B)$ is described approximately by~\cite{kokkelmans02}
\begin{equation}
  a(B)=a_{bg}\left(1-\frac{\Delta B}{B-B_0} \right),
\end{equation}
where $B$ is the applied field, $a_{bg}$ is the background scattering
length, and $B_0$ is the field at which the resonance occurs. The resonance width $\Delta
B$ is proportional to the strength of the coupling between the open
and close channels. For $^{40}$K, $a_{bg}=174a_0$~\cite{Markus03}, where $a_0$ is the Bohr radius. Based on this, we construct the driving terms and solve
Equations~(\ref{eq:Fs}) and (\ref{eq:Fa}). We find that the
scattering length tends to smooth out as it approaches the resonance, as
shown in Fig.~\ref{fig:graph1}. On the BCS side far from the
resonance, the many-body results give exactly two-body physics. When
the system is driven to the resonance, two-body scattering gives a
diverging unrenormalized (bare) scattering length. Many-body exchange fluctuations greatly suppress the divergence of the scattering length. In this region, the exchange
fluctuations are quite strong and act as a feedback to the system. On the Bose-Einstein condensate (BEC) side far from the resonance, the medium effects reduce the effect of the interaction. This is similar to what occurs on the BCS side far from the resonance.
The suppression of divergence close to the resonance is also suggested
in the many-body renormalized scattering cross section, as shown in
Fig.~\ref{fig:graph2}, where scattering occurs between  two hyperfine species, namely $|f=9/2,m_f=-9/2>$ and $|f=9/2,m_f=-7/2>$.
Here $f$ is the total angular momentum, and $m_f$ is the corresponding magnetic quantum number. As can be seen, away from the resonance, the medium effects are small. However, close to resonance, the medium effects strongly modify the scattering cross section.
\subsection{Close to a $p$-wave resonance}
\begin{figure}
\centering
\includegraphics[scale=0.33]{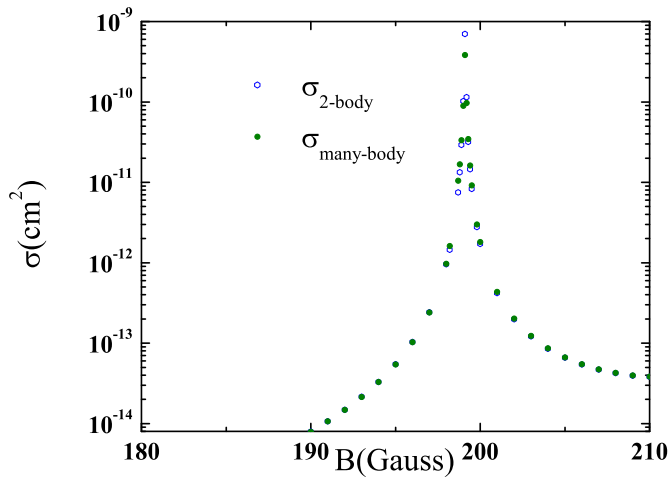}
\caption{(Color online) The scattering cross section of a $p$-wave resonance between
$^{40}K$ atom pairs in the state $|f=9/2,m_f=-9/2\rangle|f=9/2,m_f=-7/2>|\ell=1,m_{\ell}=0\rangle$ at
$T=3.2\mu K$.}\label{fig:graph3}
\end{figure}

\begin{figure}
\centering
\includegraphics[scale=0.33]{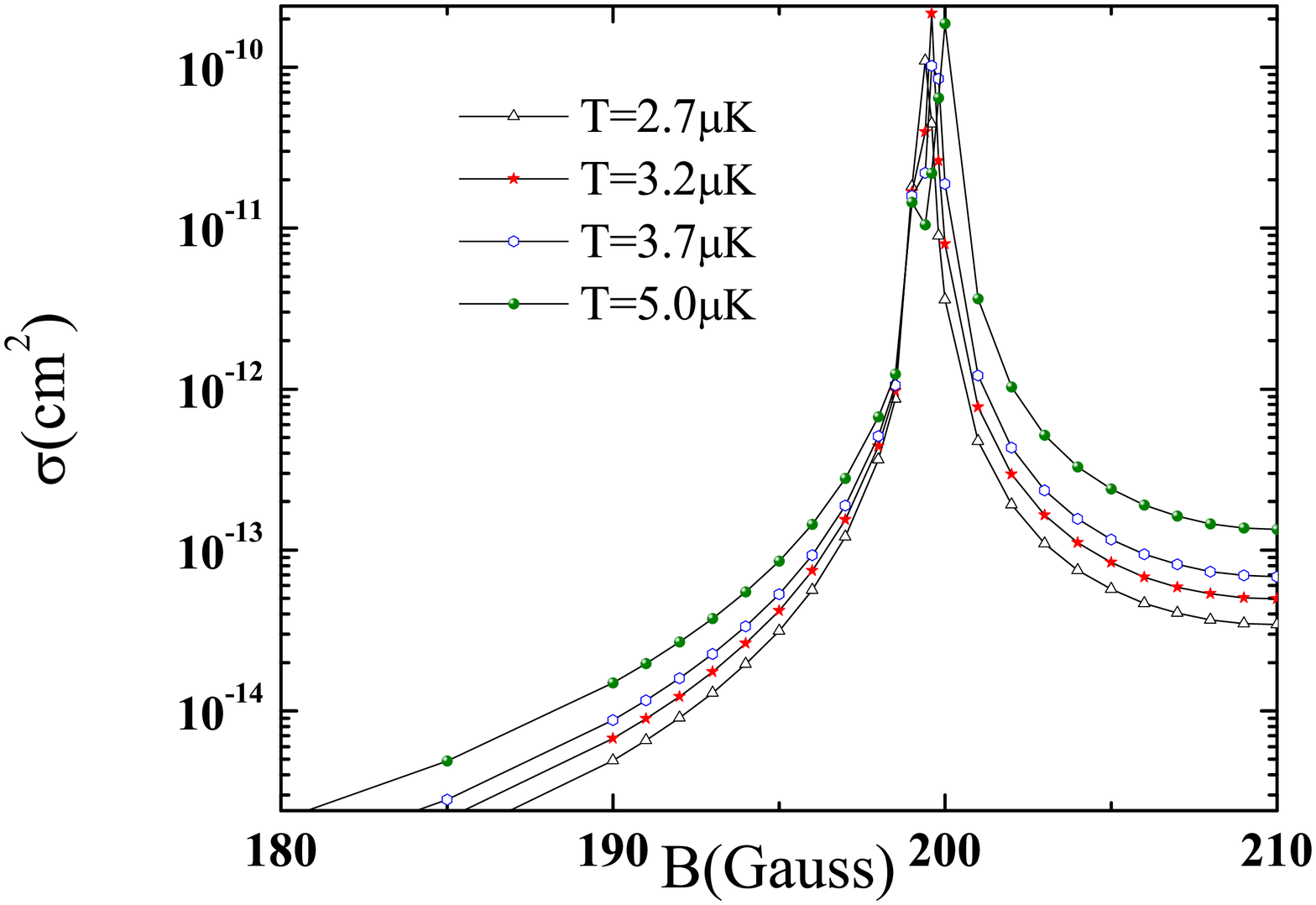}
\caption{(Color online) The many-body scattering cross section of a $p$-wave resonance for $^{40}$K atom pairs
in the $|f=9/2,m_f=-9/2\rangle|f=9/2,m_f=-7/2\rangle|\ell=1,m_{\ell}=0\rangle$ state at various temperatures.
}\label{fig:graph4}
\end{figure}

\begin{figure}
\centering
\includegraphics[scale=0.33]{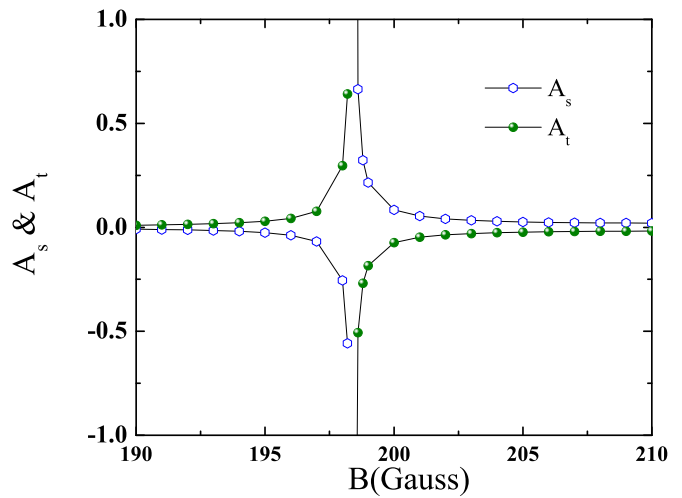}
\caption{(Color online) The singlet and triplet scattering amplitudes ($A_s$ and $A_t$) for
a $p$-wave resonance between $^{40}$K atom pairs in the state
$|f=9/2,m_f=-9/2\rangle|f=9/2,m_f=-7/2\rangle|\ell=1,m_{\ell}=0\rangle$ at
$T=0.1\mu K$.}\label{fig:graph5}
\end{figure}
\begin{figure}
\centering
{\scalebox{0.70}{\includegraphics[clip,angle=0]{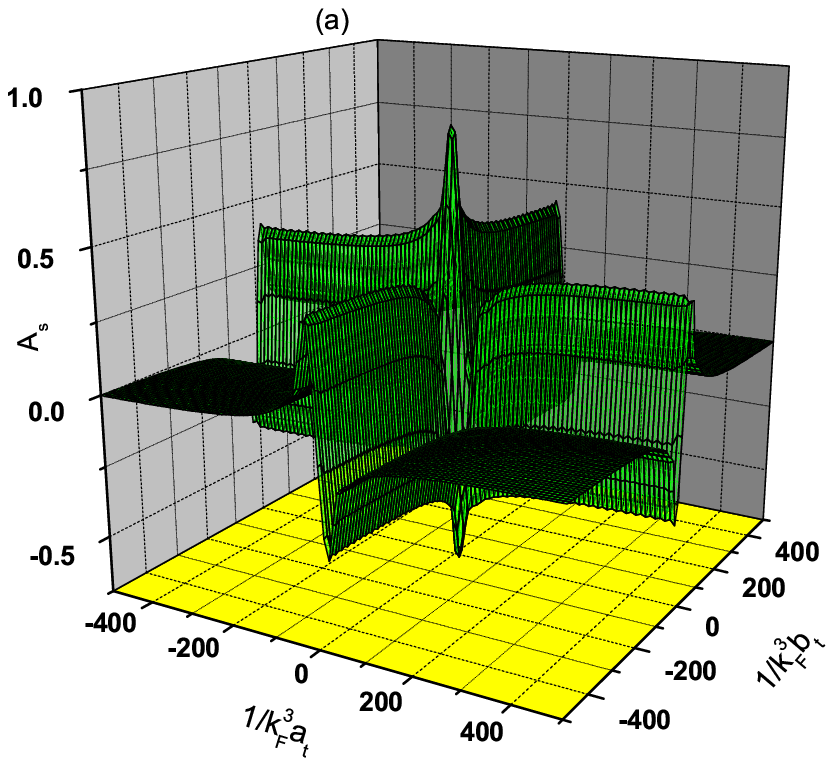}}}
{\scalebox{0.70}{\includegraphics[clip,angle=0]{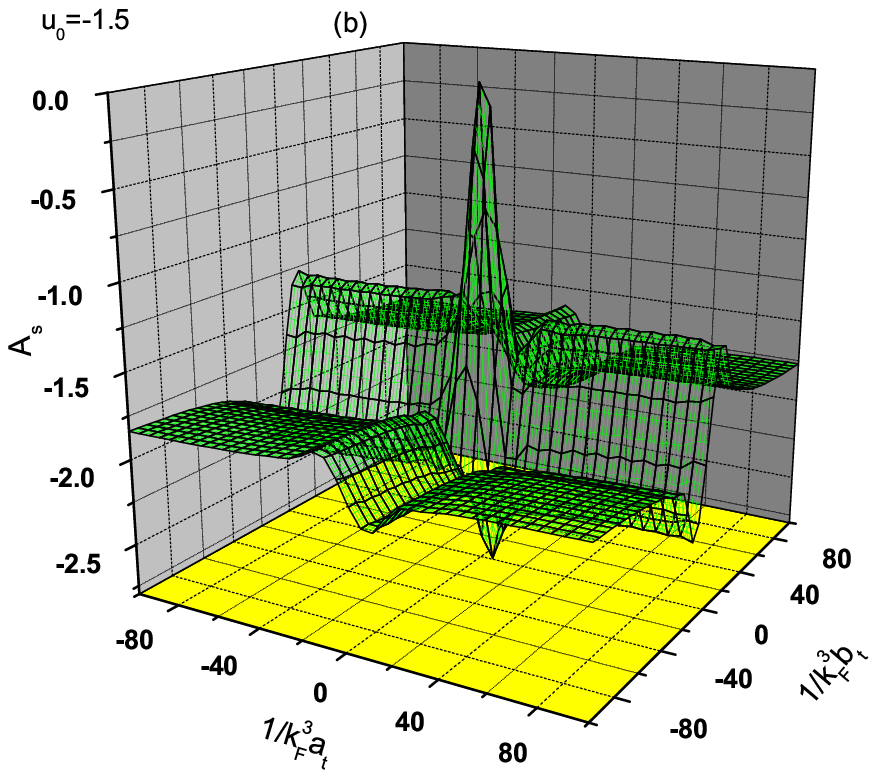}}}
\caption{(Color online) The singlet scattering amplitude $A_s$ for a $p$-wave resonance:
(a) without $s$-wave background and (b) with $s$-wave background where $U_0=-1.5$.}\label{fig:graph6}
\end{figure}

\begin{figure}
{\scalebox{0.48}{\includegraphics[clip,angle=0]{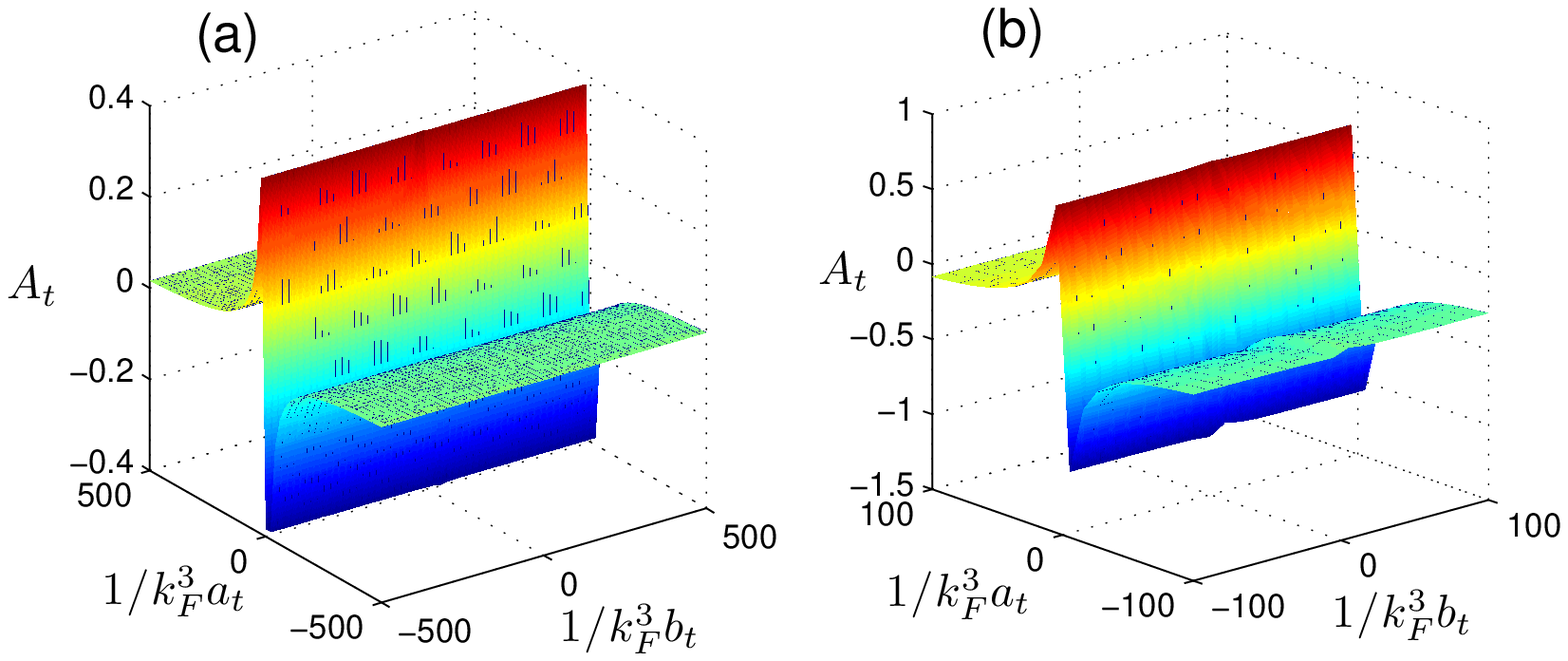}}}

\caption{(Color online) The triplet scattering amplitude $A_t$ for a $p$-wave resonance:
(a) without $s$-wave background; (b) with $s$-wave background where $U_0=-1.5$.}\label{fig:graph7}
\end{figure}
A $p$-wave resonance is distinct from an $s$-wave ($l=0$) resonance in
that the atoms must overcome a centrifugal barrier to couple to the
bound state. It is sensitive to the temperature and the magnetic field. We
need two parameters to characterize the $p$-wave resonance: scattering
volume $a_t$ and effective range $r_1$. The magnetic field
dependence of $a_t$ and $r_1$ is obtained from the fitting formula
given by Ticknor {\it{et al.}}~\cite{Ticknor04}. Therefore we can construct the driving
terms and solve equations~(\ref{eq:Fs}) and (\ref{eq:Fa}). The $p$-wave resonance could be tuned by the magnetic field to occur between two atoms in the hyperfine states of $|f,m_f>=|9/2,-9/2\rangle$ and $|f,m_f>=|9/2,-7/2\rangle$. The joint state of the atom pair will be written as $|f_1m_{f_1}\rangle|f_2m_{f_2}\rangle|\ell m_{\ell}\rangle$. The
many-body effects in a $p$ wave are less severe than in an $s$ wave, as shown
in Fig.~\ref{fig:graph3}. In Fig.~\ref{fig:graph3}, we plot the scattering cross section for a $p$-wave resonance for $^{40}$K atom pairs in the state $|f=9/2,m_f=-7/2\rangle|\ell=1,m_{\ell}=0\rangle$. Here $\ell=1$ is the orbital angular momentum quantum number, and $m_{\ell}=0$ is the corresponding magnetic quantum number. Away from the resonance, the many-body corrections due to medium effects are small. Distinct correction only appears at a
region close to the Feshbach resonance. Figure~\ref{fig:graph4} shows the
temperature dependence of the many-body scattering cross section. We can see that the
lower the temperature is, the higher the resonance peak is. As the
temperature rises, the resonance cross section broadens. The position of the resonance changes slightly with temperature due to the temperature dependence of the scattering cross section. The resonance at $T=5.0\mu$K (green dots) shows a double-peak feature resulting from the strong energy dependence of the cross section. The pairing-channel scattering amplitudes ($A_s$ and $A_t$) are shown in
Fig.~\ref{fig:graph5}. The triplet scattering amplitude $A_t$ is
negative on the BCS side, indicating a $p$-wave superfluid pairing. On the BEC side, the
singlet scattering amplitude is negative, indicating the formation of $s$-wave
molecules. The $p$-wave Feshbach resonances offer a means to experimentally study anisotropic interactions in systems other than identical fermions. On resonance the $p$-wave cross section becomes comparable to the background $s$-wave scattering. This means that it could have an equally important role in determining the collisional behavior and mean-field interaction of the quantum gases. Utilizing $p$-wave Feshbach resonances, the Joint Institute for Laboratory Astrophysics (JILA) group has successfully created $p$-wave molecules~\cite{GAE07}.
\subsection{Close to a $p$-wave resonance with an $s$-wave background}

For a $p$-wave resonance, the JILA group~\cite{Regal032} found that
there exists non-negligible off-resonant scattering in the ultracold Fermi gas of $^{40}$K. Our model can
be employed to study the situation when the singlet correlation ($s$-wave scattering) and
the triplet correlation ($p$-wave scattering) are both present. This situation can occur when the resonant magnetic field for $p$ waves and $s$ waves are well separated. By tuning the magnetic field around $p$-wave resonance one can realize a $p$-wave resonance with an $s$-wave background. The driving terms can then be modeled as
\begin{eqnarray}
   D_0^s&=&\frac{3\pi}{m}k_F^2(2a_t+b_t)N(0)+\frac{2\pi}{m}a_sN(0)\nonumber,\\
   D_0^a&=&\frac{3\pi}{m}k_F^2(2a_t-b_t)N(0)-\frac{2\pi}{m}a_sN(0)\nonumber,\\
   D_1^s&=&-\frac{3\pi}{m}k_F^2(2a_t+b_t)N(0)\nonumber,\\
   D_1^a&=&-\frac{3\pi}{m}k_F^2(2a_t-b_t)N(0).
\end{eqnarray}
Here $s$-wave background interaction is characterized by $U_0=\frac{4\pi}{m}a_sN(0)$. By fixing $U_0$
and varying the parameters $1/k_F^3a_t$ and $1/k_F^3b_t$, we can evaluate the pairing channel
scattering amplitudes ($A_s$ and $A_t$) in full parameters space through solving crossing-symmetric equations~(\ref{eq:Fs}) and (\ref{eq:Fa}). From the obtained Landau parameters $F_l^{s,a}$, by $s$-$p$ approximation~\cite{PAT75}, one can straightforwardly construct pairing-channel scattering
amplitudes via
\begin{eqnarray}
  A_{singlet}&=&A_0^s-3A_0^a-A_1^s+3A_1^a,\nonumber\\
  A_{triplet}&=&A_0^s+A_0^a-A_1^s-A_1^a.
\end{eqnarray}
The calculated singlet
scattering amplitude is plotted in Fig.~\ref{fig:graph6}. When
there is no background $s$-wave scattering, the singlet scattering
amplitude becomes very large when it crosses the unitary limit (strongly interacting regime) of either
parameter $1/k_F^3a_t$ or $1/k_F^3b_t$, as suggested in Fig.~\ref{fig:graph6}$(a)$. To investigate the effects of the $s$-wave background on the $p$-wave resonance, we choose $U_0=-1.5$ for illustration such that it has a noticeable effect. In the presence of background
$s$-wave scattering, the singlet scattering amplitude becomes negative and shows a pronounced peak when it
crosses the unitary limit of both parameters $1/k_F^3a_t$ and
$1/k_F^3b_t$. Away from the unitarity limit, the singlet scattering amplitude $A_s$ is
mainly negative, manifesting trends of  background $s$-wave pairing. The purpose of
introducing $U_0$ is to introduce background singlet pairing interaction.
The competition between the $p$-wave scattering and background $s$-wave scattering
determines the sign of the singlet scattering amplitude. When the $p$-wave interaction is weak,
the properties of the system are dominated by $s$-wave behavior with a constant negative
singlet pairing amplitude, as can be seen in the Fig.~\ref{fig:graph6}$(b)$.

To investigate $p$-wave pairing, we plot the triplet scattering amplitude $A_t$ in
Fig.~\ref{fig:graph7}. Interestingly, the triplet scattering amplitude $A_t$ is sensitive
to parameter $1/k_F^3a_t$ but insensitive to parameter $1/k_F^3b_t$ except
building up a bump near the unitarity regime in the presence of background $s$-wave scattering. Remarkably, in the absence of background $s$-wave scattering $A_s$ is antisymmetric with
respect to parameter $1/k_F^3a_t$ while symmetric with respect to parameter
 $1/k_F^3b_t$. These properties are closely related to the parametrization and symmetrical features in Equation (28), where $s$-wave scattering only contributes to the $\ell=0$ channel in the driving terms. The background $s$-wave
scattering shows its existence by shifting the triplet scattering amplitude in total
toward the negative side. In addition, it enhances the effect of parameter  $1/k_F^3b_t$,
especially near resonance. The antisymmetry of $A_t$ with respect to parameter $1/k_F^3a_t$
is lost; in contrast, the symmetry of $A_t$ with respect to parameter $1/k_F^3b_t$ is still preserved.
It is necessary to point out that when both $A_s$ and $A_t$ are negative, the actual pairing
symmetry depends on their relative magnitude.
\section{Concluding Remarks}
By using the crossing-symmetric method to treat many-body Fermi systems, one generally
solves nonlinear coupled crossing-symmetric equations for four-point vertex functions
in particle-particle, particle-hole, and exchange particle-hole channels. In appropriate limits
on the Fermi surfaces, these vertex functions become the Landau quasiparticle interaction $F(q)$
and the scattering amplitude $A(q)$. For isotropic systems, these can be expressed in Legendre polynomials giving Landau parameters $F_l^s$ and $F_l^a$ in spin-symmetric ($s$) and spin-antisymmetric ($a$) channels. From the obtained Landau parameters, one can calculate several
thermodynamic, transport, and pairing properties of a system.\\

We study $s$- and $p$-wave Feshbach resonance with the crossing-symmetric method.
Our findings are as follows: (1) many-body exchange effects may be important
close to a Feshbach resonance. Renormalized physical quantities get
smoothed out at the Feshbach resonance. In particular, we find that the particle-hole exchange fluctuations introduce an effective scattering length which has been substantially reduced close to resonances. (2) For a $p$-wave resonance, the triplet
scattering amplitude is negative on the BCS side, indicating
a triplet($p$-wave) superfluid pairing. (3) Background off-resonant scattering has
some effects on the singlet and triplet scattering amplitudes, which may
influence the pairing symmetry of the ground state.

\begin{acknowledgments}
 We are grateful for stimulating discussions with K. S. Bedell and S. Gaudio. This work was supported by the NSFC under Grants No. 10934010 and No. 60978019, the NKBRSFC under Grants No. 2009CB930701, No. 2010CB922904, No. 2011CB921502, and No. 2012CB821300, and the NSFC-RGC under Grants No. 11061160490 and No. 1386-N-HKU748/10.
\end{acknowledgments}
%

\end{document}